 \documentstyle[12pt,epsfig]{article}
\begin{document}
\begin{flushright}
CERN-TH/97-297 \\
\end{flushright}
\vspace{0.5cm}
\begin{center}
{\LARGE \bf Target Fragmentation at Polarized HERA:  \\ A Test of Universal 
Topological Charge Screening in QCD \footnote{\normalsize {Contribution to   the proceedings of the 1997 Workshop on `Physics with Polarized Protons at HERA', DESY-Hamburg, DESY-Zeuthen and CERN,  March--September 1997.}}}

\vspace{1cm}
{\Large D. de Florian $^a$,  G.M. Shore $^b$, G. Veneziano $^a$ }

\vspace*{1cm}
{\it $^a$Theoretical Physics Division, CERN, CH 1211 Geneva 23, Switzerland \\ 
$^b$ Department of Physics, University of Wales Swansea,\\ Singleton Park, 
Swansea SA2 8PP, U.K.}
\vspace*{1cm}
\end{center}
\begin{abstract}
 Topological charge screening has been proposed as the mechanism responsible
for the anomalous suppression in the first moment of the polarized proton
structure function -- the `proton spin' effect. An immediate consequence is
that this suppression should be target-independent, since the
screening is a fundamental property of the QCD vacuum. Here, we study the
possibility of testing the target-independent suppression in semi-inclusive
target fragmentation processes at polarized HERA.
\end{abstract}
\vspace{0.4cm}
\begin{flushleft}
CERN-TH/97-297 \\
November 1997
\end{flushleft}
\vspace{0.8cm}
\newpage
\section{Polarized Structure Functions and Universal Screening of
Topological Charge}

The last decade has seen an important advance in our understanding of polarized 
nucleon structure functions as a result of inclusive DIS data obtained by 
several experimental collaborations. It is now confirmed at the 10\% level 
that the Bjorken sum rule
for $g_1^p$ and $g_1^n$
is correct. On the other hand, the Ellis-Jaffe sum rule is violated or,  
equivalently, the flavour singlet axial charge $a^0$ of the proton or neutron 
is significantly less than its OZI value. 

In the QCD parton model, the breaking of the Ellis-Jaffe sum rule is understood
as due to a positive polarized gluon distribution and/or a negative polarized
strange quark distribution in $a^0$. The interpretation of the axial
charges in terms of polarized parton distributions depends on the factorization scheme.
In the AB scheme,
\begin{eqnarray}
a^3 = \Delta u - \Delta d \, , \,\, \,\,\,  \,\,\,\, 
a^8 = \Delta u + \Delta d - 2\Delta s  \, , \,\, \,\,\,  \,\,\,\, 
a^0(Q^2) = \Delta\Sigma - n_f {\alpha_s(Q^2)\over 2\pi} \Delta g(Q^2) \, .
\end{eqnarray}
All the scale dependence of $a^0(Q^2)$ is assigned to the polarized gluon, 
leaving the singlet quark distribution $\Delta\Sigma = \Delta u + \Delta d + 
\Delta s$ scale-invariant. The Ellis-Jaffe sum rule arises from the 
approximation $a^0(Q^2) = a^8$, equivalent to the OZI rule.
(For a recent theoretical review of the `proton spin' effect, 
see e.g.~ref. \cite{s}.)

An interesting conjecture is that the observed suppression in $a^0(Q^2)$
is due overwhelmingly to the gluon distribution $\Delta g(Q^2)$. If so, the 
strange quark distribution in the proton is almost zero, $\Delta s \simeq 0$, 
and $\Delta\Sigma \simeq a^8$ (preserving the spirit of the original Ellis-Jaffe
proposal). This is a theoretically appealing idea because it is the axial $U(1)_A$
anomaly (which is due to the gluons and is responsible for OZI violations in 
other channels) that is responsible for the scale dependence of $a^0(Q^2)$ and
$\Delta g(Q^2)$, whereas $\Delta\Sigma$ is scale-invariant.

However, from the present inclusive proton and neutron data, it is not possible
to isolate the quark distributions for each flavour, while information on the
polarized gluon distribution comes only from analysing the scale dependence
of $g_1^N(Q^2)$. It is therefore not yet possible to determine accurately the
contributions of $\Delta s$ and $\Delta g$ to the violation of the Ellis-Jaffe
sum rule.

To make further progress in understanding the polarized structure of the nucleon,
it is absolutely necessary to study less inclusive processes. The usual
semi-inclusive asymmetries will be essential for determining the quark flavour
decomposition, whereas open charm, jet and charged hadron production (for both
DIS and real photons) will give information on the polarized gluon distribution.
All of these are {\it current fragmentation} processes and most can be studied
at polarized HERA. However, in contrast to fixed-target experiments such as
COMPASS, polarized HERA is unique in being able to study the {\it target
fragmentation} region. We now explain the importance of target fragmentation data
in revealing the ultimate origin of the OZI breaking responsible for the violation
of the Ellis-Jaffe sum rule.

It has been proposed that the anomalous suppression of the first moment of the 
polarized structure function is not a special property of the proton and neutron
or of their parton distributions, but is a {\it target-independent} phenomenon,  which
would hold for any hadron \cite{nsv}. It is due to a universal screening of 
topological charge inherent in QCD itself.

To understand this screening mechanism, note first that using the axial $U(1)_A$
anomaly $\partial^\mu A_\mu^0 = 2n_f {\alpha_s\over8\pi} G\tilde G$, the flavour
singlet axial charge is just the forward matrix element of the gluon topological
charge density ${\alpha_s\over8\pi}G\tilde G$, i.e.
\begin{equation}
a^0(Q^2) \,\overline{N} \gamma_5 N= {1\over2M}2n_f \langle N|{\alpha_s\over8\pi}G\tilde G|N\rangle \, .
\end{equation}
The matrix element is then decomposed  into a composite operator propagator
and a vertex, which is chosen to be scale-independent. It is conjectured in 
refs. \cite{nsv,sv} that all the OZI violation in $a^0(Q^2)$ resides in the scale-dependent
propagator, while the vertex, which contains all the information on the target,
is well approximated by the OZI rule. This implies
\begin{equation}
a^0(Q^2) = s(Q^2)\, a^0_{OZI} \, ,
\end{equation}
where, for the proton or neutron, $a^0_{OZI} = a^8$ and $s(Q^2)$ is a universal 
suppression factor. Using chiral Ward identities, $s(Q^2)$ can be shown to be 
determined by the QCD topological susceptibility $\chi(k^2)$, which measures the 
response of the QCD vacuum to topological charge. Precisely,
\begin{equation}
s(Q^2) = {\sqrt{2n_f}\over f_\pi} \sqrt{\chi'(0)} \, ,
\end{equation}
where $\chi'(k^2)=d\chi(k^2)/dk^2$ and
\begin{equation}
\chi(k^2) =  \int dx \, e^{ik.x} i \langle 0|T~{\alpha_s\over8\pi}G\tilde G(x) ~
{\alpha_s\over8\pi}G\tilde G(0)|0\rangle \, .
\end{equation}
The existing inclusive proton and neutron data suggest that
$s$ is in the range 0.3 to 0.7 at $Q^2=10$ GeV$^2$,
while a QCD spectral sum rule calculation \cite{nsv} in the chiral limit
gives $s \simeq 0.6$.

The physical picture is therefore that the QCD vacuum screens the topological 
charge in the matrix element for $a^0(Q^2)$, and so the singlet axial charge of 
any hadron is OZI-suppressed by a universal factor $s(Q^2)$. The mechanism is 
analogous to the screening of electric charge in QED. There, because of the Ward 
identity, the screening is given entirely by the (`target-independent') 
dressing of the photon propagator by the vacuum polarization diagrams 
(cf. eqs. (4),(5)), leading to the relation $e_R= e_0 \sqrt{Z_3}$
(with $Z_3 < 1$) between the renormalized and bare charges, 
in close analogy with eq. (3) above.

To test this picture directly we would need to perform DIS experiments on  
targets other than the proton and neutron. The proposal of ref. \cite{sv} is that 
this can in effect be done by studying semi-inclusive processes in which a single
hadron carrying a large target energy fraction is detected in the target 
fragmentation region.

\vskip0.4cm
\centerline{
{\epsfxsize=4.0cm\epsfbox{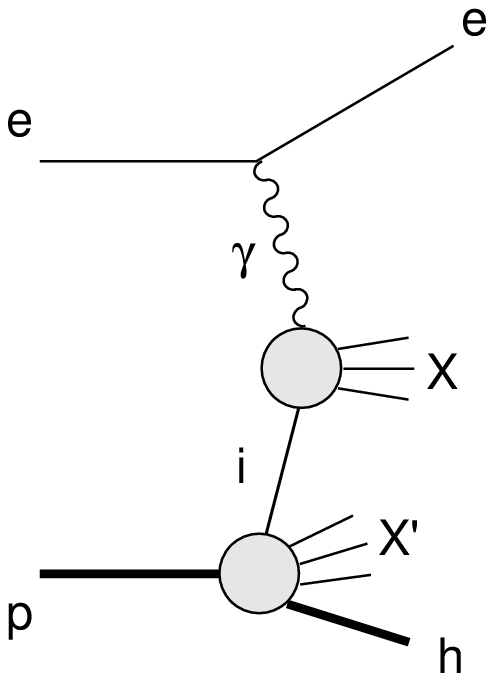}}}
\centerline{ Fig.1 ~~Semi-inclusive DIS: target fragmentation region.}
\vskip0.6cm
Semi-inclusive DIS in the target fragmentation region, as shown in Fig. 1, is best
described using (extended) fracture functions \cite{tv,gtv} 
$M_i^{h/N}(x,z,t,Q^2)$, which represent the joint probability distribution for 
producing a parton $i$ with momentum fraction $x$ and a detected hadron $h$ 
carrying an energy fraction $z=p_2'\cdot q/p_2\cdot q$ from a nucleon $N$; 
$t$ is the invariant momentum transfer. The polarized cross section 
is given at LO  \cite{dgs} (neglecting NLO  current fragmentation effects) by
\begin{equation}
\label{eq:m1}
{d\Delta\sigma^{target}\over dx dQ^2 dz dt} = \frac{4\pi\alpha^2 y(2-y)}{Q^4}   
\Delta M_1^{h/N}(x,z,t,Q^2) \, ,
\end{equation}
with the fracture equivalent of the inclusive $g_1$
\begin{equation}  
\Delta M_1^{h/N}(x,z,t,Q^2) = \sum_i \frac{\hat e_i^2}{2} 
\Delta M_i^{h/N}(x,z,t,Q^2)\, . 
\end{equation}

\vskip0.4cm
\centerline{
{\epsfxsize=4.0cm\epsfbox{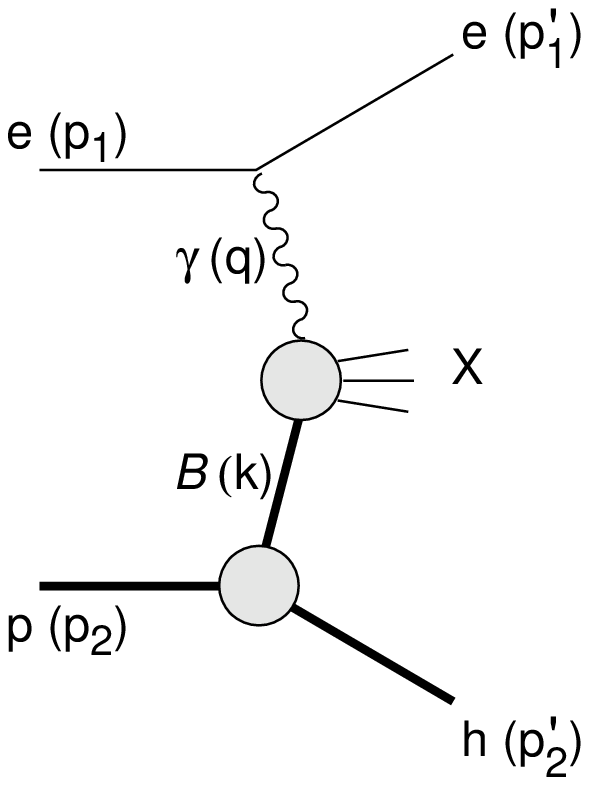}}}
\centerline{ Fig.2 ~~Single Reggeon exchange model of 
${ ep \rightarrow eh}X$.}
\vskip0.6cm
For large $z$, i.e. for the hadron carrying a large energy fraction, the process may
be simply modelled by a single Reggeon exchange diagram (Fig. 2), 
which corresponds to the approximation
\begin{equation}
\Delta M_1^{h/N}(x,z,t,Q^2)~
{\mathrel{\mathop\simeq_{z\rightarrow 1}}}~ 
F(t) (1-z)^{-2\alpha_{{\cal B}}(t)} g_1^{{\cal B}}\left( {x\over 1-z}, t, 
Q^2\right) \, ,
\end{equation}
where $g_1^{{\cal B}}$ is the structure function for the exchanged 
Reggeon ${\cal B}$ with trajectory $\alpha_{{\cal B}}(t)$. For forward events 
at HERA, $t\simeq 0$, and the fracture functions may be integrated over to increase statistics.

Although the Regge form is only an approximation to the more fundamental QCD 
description in terms of fracture functions, it shows clearly how observing 
semi-inclusive processes at large $z$, with particular choices of $h$ and $N$, 
amounts in effect to performing inclusive DIS on virtual hadronic targets 
${\cal B}$. For example, from the quark diagram in Fig. 3, we see that the
reaction ${ ep\rightarrow e}\pi^- X$ measures $g_1^{{\cal B}}$ for ${\cal B} = 
\Delta^{++}$.

\vskip0.4cm
\centerline{
{\epsfxsize=8.0cm\epsfbox{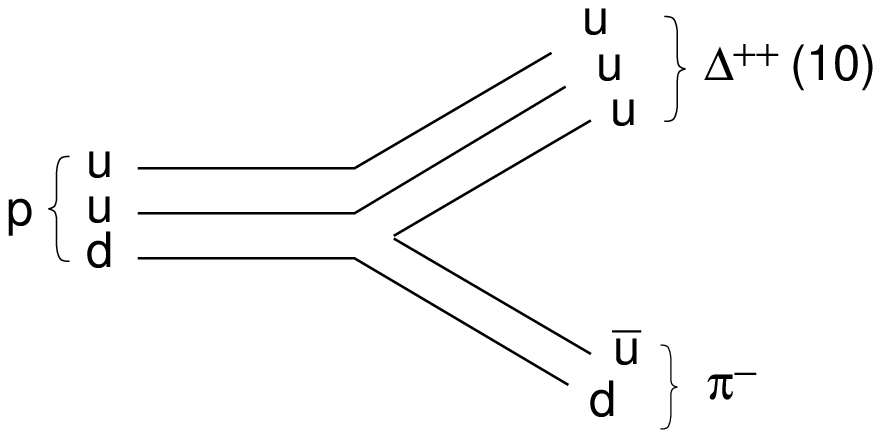}}}
\centerline{ Fig.3~~Quark diagram for the ${ Nh}{\cal B}$ vertex in
the reaction ${ ep \rightarrow e}\pi^- X$} 
\centerline{ where ${\cal B}$ has the quantum numbers of $\Delta^{++}$.}
\vskip0.6cm
Since the suppression mechanism for topological charge depends only on the
flavour structure of the exchanged object, we can make simple predictions for the
ratios ${\cal R}$ of the first moments of the polarized fracture functions 
$\int_0^{1-z} dx \, \Delta M_1^{h/N}(x,z,t,Q^2) $ for various reactions.
We emphasize that these do not depend on any detailed model of the fracture 
functions, such as single- or even multi-Reggeon exchange.
The most interesting is \footnote{In these equations, a ratio of Wilson
coefficients $C_1^{S}/C_1^{NS}$ has been absorbed into the definition 
of $s(Q^2)$.}
\begin{equation}
\label{eq:ratio}
{\cal R}\biggl({ { ep\rightarrow e} \pi^- X \over { en\rightarrow e} \pi^+ X}
\biggr) = {2s+2\over2s-1}
\end{equation}
for which we expect a clear difference from the naive quark counting (OZI)
expectation of 4.

For strange mesons, the ratio depends on whether the exchanged object has $SU(3)$
quantum numbers in the ${\bf 8}$ or ${\bf 10}$ representation, so the prediction
is less clear:
\begin{eqnarray}
{\cal R}\biggl({{ ep\rightarrow e K}^0 X \over { en\rightarrow e K}^+ X}\biggr)
&=& {2s-1-3(2s+1)F^*/D^* \over 2s-1-3(2s-1)F^*/D^*}~~~~({\bf 8}) \\
{}&=& {2s+1\over2s-1} ~~~~\,\,\,\,\,\,\,\,\,\,\,\,\,\,\,\,\,\,\,\,\,\,\,\,\,\,\,\,\,\,\,\,\,\,\,\,\,\,\,\,\,\,\,\,\,\,({\bf 10}) \, .\nonumber 
\end{eqnarray}
Similar results can be given for charmed mesons. For example
\begin{equation}
{\cal R}\biggl({{ ep\rightarrow e D}^- X \over { en\rightarrow e D}^0 X}\biggr)
= {2s+2\over2s-1}\, .
\end{equation}
For $z\rightarrow 0$, these ratios all reduce to the first moment ratio
of $g_1^p/g_1^n$. The expectation for the whole range $0<z<1$ is therefore as
shown in the sketch of Fig. 4.

\vskip0.6cm 
\centerline{
{\epsfxsize=7cm\epsfbox{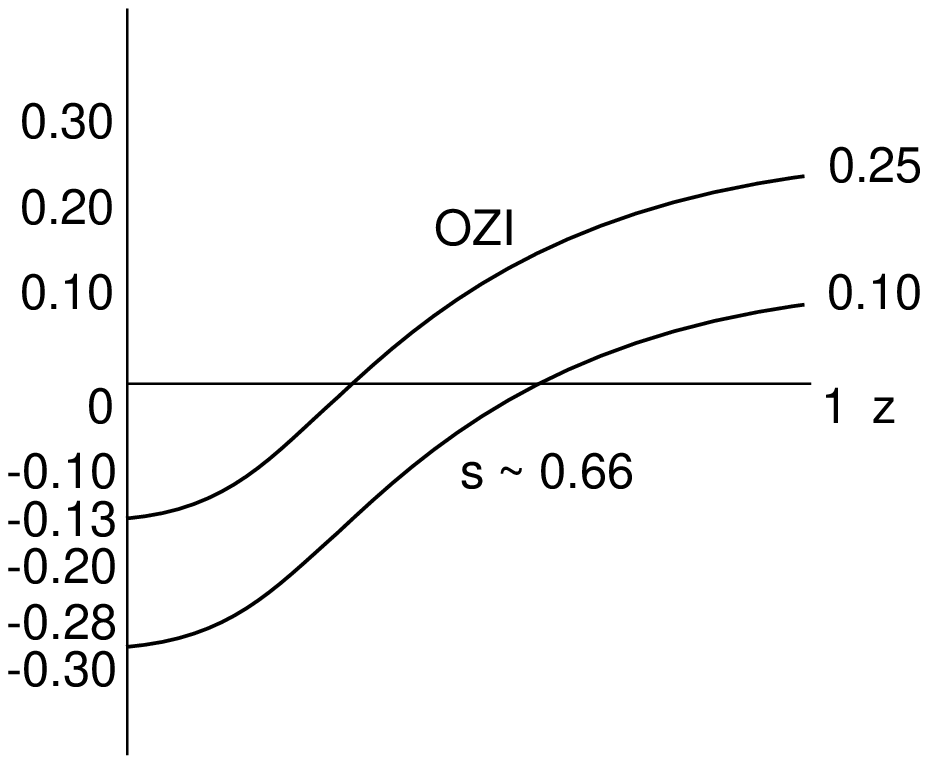}}}
\centerline{ Fig.4~~Cross-section ratios for ${ en \rightarrow e}\pi^+
({ D}^0)X$ over ${ ep \rightarrow e}\pi^-({ D}^-)X$}
\centerline{between $z\rightarrow 0$ and $z\rightarrow 1$, contrasting the OZI 
and the target-independence predictions.}
\vskip0.6cm
 

\section{Target Fragmentation at Polarized HERA}

In this section we summarize the most important experimental and phenomenological
features of a measurement in the target fragmentation region, 
concentrating on the case of pion production. 

It is worth noticing that the same features apply for $\rho$ production, since 
the quantum numbers of the exchanged object are the same, and, indeed, for any 
non-strange meson with the same charge. Since the production of leading strange 
mesons from protons and neutrons is highly suppressed, the particle identification
requirements will be less stringent than for pions alone. 
It is also important to emphasize that the predictions given here require 
dominance by the exchange of an object in the ${\bf 10}$ representation of 
flavour $SU(3)$ over one with the same quantum numbers in the {\bf 27}, 
corresponding to disconnected quark diagrams (cf. ~Fig. 3) with 5-quark
exchange. Assuming this, the predictions are expected to apply not only for values 
of $z$ very close to 1, but for a more relaxed kinematical region of leading 
mesons, say $z>0.6$. In any case, the $z$ dependence of the ratio will also give 
very useful information about this mechanism.

In order to clarify the experimental situation it is useful to compare it with 
the well-known diffractive and leading protons and neutrons measurements already 
made by both HERA collaborations.
Leading neutron production was measured by the ZEUS \cite{zeusn} and H1 \cite{h1n} 
collaborations using a Forward Neutron Calorimeter (FNC), a lead scintillator 
calorimeter located about 100 m downstream of the interaction point, allowing the 
detection of very forward scattered neutral particles within about 1 mrad of the 
incident proton beam direction. Both collaborations found that about 10\% of the 
DIS events contain a neutron in the forward region with more than half of the 
energy of the proton beam.

In the case of protons in the target fragmentation region, they are measured with
the Leading Proton Spectrometer (LPS), consisting of detectors operating inside 
movable Roman pots located in the beam pipe, which measure the curvature of the 
charged particle and therefore its momentum. The ZEUS Collaboration \cite{zeusp} 
has measured the distribution of leading protons in the range of $z>0.6$, 
analysing also the $t$ dependence. This allowed a cleaner definition of the 
diffractive process in terms of fracture functions, i.e.~as one where a proton 
carrying more than 97\% of the momentum of the initial proton is detected in the final 
state, instead of the usual definition in terms of rapidity gaps. The H1 
Collaboration \cite{h1p} has measured the production of leading protons in the 
kinematical region of $0.7<z<0.9$ with enough accuracy to analyse the $Q^2$ 
dependence. The sample of data used corresponds to an integrated luminosity 
of about $1.44$ pb$^{-1}$, to be compared with the maximum one expected in the 
polarized case of about $500$ pb$^{-1}$.

The variables used in the different measurements are easily related to $z$,   
by $z=x_L$ for leading protons and $z=1-x_{I\hspace{-0.2em}P}$ for diffractive 
scattering. In the same way, the defined leading proton, leading neutron and 
diffractive structure functions are just $M_2^{B/p}(x,z,Q^2) = 
\sum_i \hat{e}_i^2 x M_i^{B/p}(x,z,Q^2)$, the fracture function equivalent 
of the inclusive $F_2$, where $B$ corresponds to either a proton or a neutron.

The acceptance of the LPS has been optimized for positively charged particles 
with an almost negligible one for negative hadrons, since proton measurement 
was  the main objective and, up to now, there is no identification system 
available in the target region. An ID system would allow a precise measurement 
of charged mesons as needed in this proposal.
 
The acceptance of the LPS could   in principle be extended in order to measure 
mainly negative hadrons by adding some extra stations.
In that case it would be possible to obtain a very 
good approximation for the production of $M^-=\pi^-,\rho^-$ by the sensible 
assumption of its dominance over strange mesons and heavier hadrons, i.e. 
$\sigma(p\rightarrow h^-)\simeq \sigma(p\rightarrow M^-)$. Any remaining 
positive charged hadron contribution could be subtracted if the acceptance 
for both negative and positive hadrons were known.

The identification of positive mesons, needed for the process 
$n\rightarrow M^+$, seems to be the most complicated task, basically because 
the cross section will be dominated by the proton background.
 
Besides this,   an additional problem comes from the fact that neutron beams 
are not available. They are fundamental in the  construction of the ratio in 
eq. (\ref{eq:ratio}), which is independent of the unknown matrix elements 
appearing in each sum rule separately. The polarized HERA option for neutrons 
is  $^3$He. In that case, even for the polarized cross section, there is a 
contribution coming from the scattering of protons in  $^3$He, which has  
to be subtracted. Assuming for the sake of simplicity that $^3$He is a 
simple combination of free protons and neutrons, i.e.~schematically 
$^3$He=$ A \, p + B \, n$, the cross section for the production of 
positive hadrons measured in the LPS is given by
\begin{equation}
\sigma(^3{ {\rm He}}\rightarrow h^+)\simeq A\, \sigma(p\rightarrow h^+) + B\, 
\sigma(n\rightarrow p)+ B\, \sigma(n\rightarrow M^+)\, ,
\end{equation}
where the first contribution on the right-hand side can be obtained from 
measurements with the proton beam. In order to subtract the second one, 
it would be necessary at least to distinguish protons from mesons. 
In the region of very small $x$, where the singlet contribution dominates, 
it could be possible to use $SU(2)$ arguments to relate the cross section 
for the process $p\rightarrow n$, obtained with the existing FNC, to  
the corresponding one for $n\rightarrow p$, but the same argument fails 
for the non-singlet contribution.

It is possible to obtain a rough estimate of the fraction of DIS events 
containing one of these leading mesons by using a sensible model for the 
production of forward hadrons, as in the approach of ref. \cite{holtmann}. 
The idea there is to exploit the non-perturbative meson--baryon $(MB)$ Fock  
components of the nucleon to compute the flux of the final state particle 
in the proton. This flux is normalized by the non-perturbative strong couplings 
known from low energy physics ($g^2_{pMB}/4\pi$) and with a parameter adjusted 
to describe the distribution of the produced $B$ in hadron--hadron collisions.

In this approach, an expression for $M_2^{M/p}(x,z,Q^2)$ \cite{ds} can be 
written down: 
\begin{equation}
M_2^{M/p}(x,z,Q^2) = \phi_{M/p}(z) F_2^{B} \left( \frac{x}{1-z},Q^2 \right)\, ,
\end{equation}
i.e.~the flux (in this case integrated over $t$)
times the structure function of the exchanged object.
 
In ref. \cite{holtmann} the flux for $\phi_{ \Delta^{++}/p}(z)$ was computed 
taking into account the exchange of both $\pi$ and $\rho$. The flux 
needed in our case can be obtained from this by a simple crossing 
relation $\phi_{M/p}(z) = \phi_{B/p}(1-z)$, where $B=\Delta^{++}$. 
It turns out from the computation that at large $z$
the production of $\rho^-$ is expected to dominate over  
that of $\pi^-$, basically because a heavy meson will carry a larger 
fraction of the momentum of the $MB$ state than the lighter one. 
In contrast to ref. \cite{holtmann}, where the reggeization of the flux 
is not considered, we have also imposed an asymptotic behaviour of 
$(1-z)^{1-2\alpha_B(t)}$ with $\alpha_{\Delta}(0)\approx 0.0$,  
but its consequence is negligible since the flux in \cite{holtmann} 
already vanishes when $z\rightarrow 1$. 

The structure function of the exchanged 
virtual $\Delta^{++}$ can be estimated using parton distributions in the 
proton \cite{grv}. Even though this approach does not have the same validity as 
in the case of an almost real pion exchange, such a model 
could also be useful in order to obtain sensible extrapolations of the fracture 
function in the unmeasured region, since some of 
the parameters of the model could be fitted to reproduce the results in the measured region. 

Then the fraction of DIS events with either a $\pi^-$ or a $\rho^-$ in the 
target fragmentation region with $z>z_{min}$ is given by 
\begin{equation}
\frac{d\sigma(ep\rightarrow e'M X) / dx\, dQ^2}
{d\sigma(ep\rightarrow e' X) / dx\, dQ^2} =
\frac{\int_{z_{min}}^{1-x}dz\,   M_2^{M/p}(x,z,Q^2) }{F_2^p(x,Q^2)} \, .
\end{equation}

This approach has been shown to work rather well for leading neutron production 
\cite{ds}, but to underestimate the production of leading protons by almost 
a factor of 2 \cite{h1p}.  
As a final result it is found that a fraction of between $0.5\%$ and $1\%$ 
of the DIS events will contain a leading meson in the target fragmentation 
region where the LPS has non-vanishing acceptance  ($z>0.6$) and in the 
dominant $x<0.1$ domain. Most of them, as expected, correspond to the 
lower values of $z$ and the ratio decreases for larger $x$. This prediction for the rate of meson production can be tested by using the unpolarized proton beam.

The experimental analysis of the semi-inclusive polarized cross section can be done in 
the same way as for the inclusive one containing  $g_1$,
since the expressions for the cross sections (as in (\ref{eq:m1}))
corresponding to structure and fracture functions are 
completely equivalent. In this case, a sub-sample of the DIS events, 
where a leading particle with a fraction of momentum $z$ is detected  
has to be used for the analysis. For leading protons and neutrons 
the corresponding sample of events showed the same rapidity gap features 
as the full sample; the same can thus be expected for this measurement, 
in contrast to the diffractive case where large rapidity gaps are observed.

\vskip0.6cm 

\section{Summary and Conclusions}

 We have described experiments that would 
test the idea of universal topological charge screening as the mechanism
underlying the `proton spin' effect. Obviously, many of the experimental
requirements described above involve difficult technical problems, which
remain to be solved. In particular, the identification of mesons in the LPS region seems to be the most complicated task, together with the extension of the LPS acceptance in order to measure negative particles. Furthermore, the proposed observable includes  the production of positively charged mesons   from a polarized neutron beam, which could be obtained from the polarized $^3$He measurement.  

Nevertheless,  target fragmentation studies can provide
a key to understanding some very fundamental aspects of QCD, and would
be a valuable part of the experimental programme at polarized HERA, which, we want to stress, would be the only experimental   facility available to study these particular processes.

\vskip0.2cm
We warmly thank A. de Roeck for many interesting comments and discussions.
The work of  GMS is partially supported by the EC TMR Network Grant FMRX-CT96-0008 and the one of DdF by the World Laboratory.


 \end{document}